\documentstyle[aps,version2]{revtex}
\begin{document}

\bibliographystyle{unsrt}
\def\Journal#1#2#3#4{{#1} {\bf #2}, #3 (#4)}
\def\NCA{ Nuovo Cimento}
\def\NIM{ Nucl. Instrum. Methods}
\def\NIMA{{ Nucl. Instrum. Methods} A}
\def\NP{{ Nucl. Phys.} }
\def\PLB{{ Phys. Lett.}  B}
\def\PRL{ Phys. Rev. Lett.}
\def\PRD{{ Phys. Rev.} D}
\def\ZPC{{Z. Phys.} C}
\def\ZPA{{Z. Phys.} A}
\def\st{\scriptstyle}
\def\sst{\scriptscriptstyle}
\def\mco{\multicolumn}
\def\epp{\epsilon^{\prime}}
\def\vep{\varepsilon}
\def\ra{\rightarrow}
\def\ppg{\pi^+\pi^-\gamma}
\def\vp{{\bf p}}
\def\ko{K^0}
\def\kb{\bar{K^0}}
\def\al{\alpha}
\def\ab{\bar{\alpha}}
\def\be{\begin{equation}}
\def\ee{\end{equation}}
\def\bea{\begin{eqnarray}}
\def\eea{\end{eqnarray}}
\def\CPbar{\hbox{{\rm CP}\hskip-1.80em{/}}}
\def\ua{\uparrow}
\def\da{\downarrow}

\begin{titlepage} 

\bigskip
\begin{title}
Large corrections to asymptotic
$F_{\eta_c \gamma}$ and $F_{\eta_b \gamma}$
in the light-cone perturbative QCD
\end{title}
\author{Fuguang Cao and Tao Huang}
\begin{instit}
CCAST (World Laboratory), P.O. Box 8730, Beijing 100080, P.~R.~China
\end{instit}
\begin{instit}
Institute of High Energy Physics, Academia Sinica,
P.O. Box 918, Beijing 100039, P.~R.~China
\end{instit}
\begin{instit}
and Institute of Theoretical Physics, Academia Sinica,
P.O. Box 2735, Beijing 100080, P.~R.~China\thanks{Mailing address.
E-mail address: caofg@itp.ac.cn.}
\end{instit}

\begin{abstract}
The large-$Q^2$ behavior of $\eta_c$-$\gamma$ and $\eta_b$-$\gamma$
transition form factors,
$F_{\eta_c\gamma}(Q^2)$ and $F_{\eta_b\gamma}(Q^2)$
are analyzed in the framework of light-cone perturbative QCD
with the heavy quark ($c$ and $b$) mass effect,
the parton's transverse momentum dependence
and the higher helicity components in
the light-cone wave function are respected.
It is pointed out that the quark mass effect
brings significant modifications to the asymptotic predictions
of the transition form factors 
in a rather broad energy region,
and this modification is much severer for
$F_{\eta_b\gamma}(Q^2)$ than that for $F_{\eta_c\gamma}(Q^2)$
due to the $b$-quark being heavier than the $c$-quark.
The parton's transverse momentum and the higher helicity components
are another two factors which decrease the perturbative predictions.
For the transition form factor $F_{\eta_c\gamma}(Q^2)$,
they bring sizable corrections
in the present experimentally accessible energy region
($Q^2 \leq 10 \, {\rm GeV^2}$).
For the transition form factor $F_{\eta_b\gamma}(Q^2)$,
the corrections coming from these two factors are negligible
since the $b$-quark mass is much larger than the parton's average
transverse momentum.
The coming $e^+ e^-$ collider (LEP2)
will provide the opportunity to examine these
theoretical predictions.

\bigskip
\noindent
PACS number(s): 12.38.Bx, 13.40.Gp, 14.40.Gx

\end{abstract}

\end{titlepage}

\section{Introduction}
Among a large number of exclusive processes, neutral meson production
in two-photon collision, $\gamma^* \gamma \ra P$
($P$ being $\pi^0,\eta,\eta^\prime,\eta_c,\eta_b \ ...$)
is the simplest one since two photons and one meson
are involved in the initial and final states, respectively.
Only one form factor named meson-photon transition form factor
($F_{P\gamma}$) is necessary to describe this class of processes.
Studying $F_{P\gamma}$ provides a rather simple and rigorous way to the
test of QCD and the determination of the meson wave function
(non-perturbative physics) \cite{Brodsky}.
Experimentally, a lot of collaborations
(TPC/Two-Gamma \cite{TPC}, CELLO \cite{CELLO}
CLEO \cite{CLEO} and L3 \cite{L3} {\it etc.})
have measured the form factors
$F_{\pi\gamma}(Q^2)$, $F_{\eta\gamma}(Q^2)$,
and $F_{\eta^\prime\gamma}(Q^2)$
in the $Q^2$ region up to $9,\ 20$ and $30$ GeV$^2$, respectively,
where $Q^2$ is the virtuality of the virtual photon.
Although with poor statistics, the $c\bar c$ states
($\eta_c, \chi_{c0}$ and $\chi_{c2}$) productions have been
observed \cite{Aurenche96}.
In LEP2, the dominant process is $e^+ e^-\ra e^+ e^- + X
\ (\gamma\gamma\ra X)$.
Considering the higher energy (the center of mass energy will reach $100$ GeV)
and the higher luminosity (the cross section of this process grows like
$({\rm ln} s/m^2_e)^2$ with $s$ being the invariant energy square of the
incoming $e^+ e^-$ pair, whereas the annihilation cross section decrease
like $s^{-1}$), LEP2 will be a good factory for the production of 
the heavy quarkonium production ($c\bar c$ and $b \bar b$), and
will greatly stimulate theoretical studies on these processes.
At present, it seems a measurement of $F_{\eta_c\gamma}$ up to
about $10$ GeV$^2$ is possible \cite{Aurenche96}.
Theoretically, there are also a lot of studies on these
form factors \cite{Brodsky2,Ji,Ong,Kroll,Cao,Radyushkin,Anisovich}.
In the large-$Q^2$ region, perturbative QCD can be employed as
a powerful tool.
The large-$Q^2$ behavior of form factors $F_{\pi^0\gamma}$,
$F_{\eta\gamma}$ and $F_{\eta^\prime\gamma}$ have been studied
in some detail by several
authors \cite{Brodsky2,Ji,Ong,Kroll,Cao,Radyushkin,Anisovich}.
Recently, the form factor $F_{\eta_c\gamma}$ has also been analyzed
in the convariant perturbative theory by adopting
the Breit reference frame \cite{Feldmann}.
In this note, we present a theoretical study on the $F_{\eta_c\gamma}$
and $F_{\eta_b\gamma}$
in the framework of light-cone perturbative QCD (LCPQCD).
It is pointed out that in the LCPQCD calculations of $F_{\eta_c\gamma}$
and $F_{\eta_b\gamma}$,
there are two differences from that in the case of
$F_{\pi^0\gamma}$, $F_{\eta\gamma}$ and $F_{\eta^\prime\gamma}$.
First, compared with the light quark ($u$, $d$ and $s$) masses,
the $c$- and $b$-quark masses
should not be neglected in evaluating the hard scattering amplitude,
while the quark masses involved in
the calculation of $F_{\pi\gamma}$, $F_{\eta\gamma}$
and $F_{\eta^\prime\gamma}$ can be neglected reasonably.
Second, considering the Wigner-Melosh rotation and $c$- and
$b$-quark masses being large, one finds that there are contributions
coming from the higher helicity components
in the light-cone wave functions
besides that come from the ordinary helicity components.
For the $\pi$, $\eta$ and $\eta^\prime$ mesons,
the contributions from the higher helicity components can also
be neglected in the limit of vanishing the quark masses.

\section{Light-cone formalism and light-cone wave function}

The light-cone (LC) formalism \cite{LCF} provides a
convenient framework for the relativistic description of hadrons
in terms of quark and gluon degrees of freedom,
and the application of perturbative QCD to
exclusive processes has mainly been developed in
this formalism (light-cone perturbative QCD) \cite{LCPQCD}.
In this formalism, the quantization is chosen at a particular
light-cone time $\tau=t+z$. Thereby, several characters arise
in this formalism:
i) The hadronic
wave function which describes the hadronic composite state
at a particular
$\tau$ is expressed in terms of 
a series of light-cone wave functions in Fock-state basis,
for example,
\begin{eqnarray}
| \pi \rangle=\sum | q \bar q \rangle \psi_{q {\bar q}/\pi}+
\sum | q \bar q g \rangle \psi_{q {\bar q} g/\pi}+\cdots,
\label{Fock}
\end{eqnarray}
and the temporal evolution of the state is generated by the light-cone
Hamiltonian $H_{LC}=P^-=P^0-P^3$;
ii) The vacuum is very simple.
The zero-particle state is the only one which has zero total $P^+$,
since all quanta must have positive light-cone
momentum $k_i^+$ and $P^+=\sum_i k_i^+$. The zero-particle state
can't mix with the other states which contain a certain number of particles.
Hence the vacuum state in the light-cone Fock basis (Eq. \ref{Fock}))
is an exact eigenstate of the full Hamiltonian $H_{LC}$,
and all bare quanta in a hadronic Fock state are parts of the hadron.
This point does very differ from that in the equal-$t$ perturbative theory
in which the quantization is performed at a given time $t$.
In the equal-$t$ quantization,
it is possible to make up zero-momentum state which
contains some particles, since the momentum of each particle may
be positive or negative, and the momentum of a composite state
is the sum of the momentum of each participant particle.
Thus the zero-particle state may mix with some zero-momentum
states which contain particles to build up the ground state,
which makes the vacuum become complex.
iii) The contributions coming from higher Fock states are 
suppressed by $1/Q^n$, therefore one can employ only the valence state to 
the leading order in the large-$Q^2$ region.
Light-cone perturbative QCD is very convenient for light-cone
dominated processes.
For the detail calculation rules we refer to literatures 
\cite{LCPQCD,BHL}.

The essential feature of light-cone PQCD 
applying to exclusive processes is
that the amplitudes for these processes can be written as a
convolution of hadron light-cone wave functions (LCWF)
(or quark distribution amplitudes, DA)
for each hadron involved in the process with a hard-scattering
amplitude $T_H$. Both LCWF and the $T_H$ are the basic blocks for the
LCPQCD calculation.
It has been pointed out that the Wigner-Melosh \cite{Wigner-Melosh}
rotation should be taken into account in order to 
connect the spin structures
of the light-cone wave function and that of the
instant-form wave function \cite{LCWF}.
As the Wigner-Melosh rotation is respected,
the light-cone wave function for the lowest valence state of
$\eta_c$ ($\eta_b$) can be expressed as \cite{LCWF}
\begin{eqnarray}
|\psi^{\eta_c(\eta_b)}_{q\overline{q}}>=
\psi(x,{\bf k}_{\perp},\uparrow,\downarrow)|\uparrow\downarrow>
+\psi(x,{\bf k}_{\perp},\downarrow,\uparrow)|\downarrow\uparrow>
\nonumber \\
 +\psi(x,{\bf k}_{\perp},\uparrow,\uparrow)|\uparrow\uparrow>
+\psi(x,{\bf k}_{\perp},\downarrow,\downarrow)|\downarrow\downarrow>,
\label{eq:pwf}
\end{eqnarray}
where
\begin{equation}
\psi(x,{\bf k}_{\perp},\lambda_{1},\lambda_{2})=
C^{F}_{0}(x,{\bf k}_{\perp},\lambda_{1},\lambda_{2})
\varphi(x,{\bf k}_{\perp}).
\end{equation}
Here $\varphi(x,{\bf k}_{\perp})$ is the momentum space wave function
in the light-cone formalism.
The coefficients $C_0^F(x,{\bf k}_{\perp},\lambda_{1},\lambda_{2})$
which result from the considering of the Wigner-Melosh rotation 
turn out to be \cite{LCWF}
\bea
C_0^F(x,{\bf k}_\perp,\ua,\da)&=&
\frac{m}{[2(m^2+{\bf k}_\perp^2)]^{1/2}};\nonumber\\
C_0^F(x,{\bf k}_\perp,\da,\ua)&=&
-\frac{m}{[2(m^2+{\bf k}_\perp^2)]^{1/2}}; \nonumber\\
C_0^F(x,{\bf k}_\perp,\ua,\ua)&=&
-\frac{(k_1-ik_2)}{[2(m^2+{\bf k}_\perp^2)]^{1/2}};
\label{coefficient2}\\
C_0^F(x,{\bf k}_\perp,\da,\da)&=&
-\frac{(k_1+ik_2)}{[2(m^2+{\bf k}_\perp^2)]^{1/2}}. \nonumber
\eea
where $m$ is the $c$- ($b$-) quark mass for $\eta_c$ ($\eta_b$),
and ${\bf k}_\bot$ is
the quark transverse momentum.
$C_0^F$ satisfy the relation
\bea
\sum_{\lambda_1,\lambda_2}C_0^F(x,{\bf k}_\perp,\lambda_1,\lambda_2)
C_0^F(x,{\bf k}_\perp,\lambda_1,\lambda_2)=1.
\eea
One character of the light-cone wave function is 
that there are higher helicity
($\lambda_1+\lambda_2=\pm 1$) components besides the ordinary helicity
($\lambda_1+\lambda_2=0$) components, while the instant-form
wave function has only the ordinary helicity components.
The above result means that the light-cone spin of a
composite particle is not directly the sum of its constituents'
light-cone spins but the sum of Wigner rotated light-cone spins of the
individual constituents.
A natural consequence is that in light-cone formalism a hadron's
helicity is not necessarily equal to the sum of the
quark's helicities, {\it i.e.}, $\lambda_{H}\neq \sum_{i} \lambda_{i}$.
This result has been employed in the studies of several processes:
the proton ``spin puzzle'' \cite{crisis}, proton's structure,
the ratio $F_2^n/F_2^p$, the proton, neutron, and deuteron
polarization asymmetries, $A_1^p$, $A_1^n$,
$A_1^d$ {\it etc.} \cite{others}.

\section{The meson-photon transition form factors $F_{\eta_c \gamma}$
and $F_{\eta_b \gamma}$}

In the following, we first analyze
the $\eta_c$-$\gamma$ transition form factor $F_{\eta_c \gamma}$.
The analysis for $F_{\eta_b \gamma}$ can be obtained in a similar way.
The $\eta_c$-$\gamma$ transition form factor $F_{\eta_c \gamma}$ is
extracted from the $\eta_c \gamma \gamma^\ast$ vertex,
\begin{eqnarray}
\Gamma_\mu =- i e^2 F_{\eta_c \gamma} \epsilon_{\mu \nu \alpha \beta}
p^\nu_{\eta_c} \epsilon^\alpha q^\beta,
\end{eqnarray}
where $p_{\eta_c}$ and $q$ are the momenta of the $\eta_c$ meson and the
virtual photon respectively, and $\epsilon$ is the polarization
vector of the on-shell photon.
In the standard ``infinite-momentum" frame \cite{Brodsky},
the momentum assignment can been written as
\begin{eqnarray}
p_{\eta_c}&=&(p^+,p^-,p_\bot)=(1,m_{\eta_c}^2,0_\bot),\nonumber \\
q&=&(0,q_\bot^2-m_{\eta_c}^2,q_\bot), \\
q^\prime&=&(1,q_\bot^2,q_\bot), \nonumber
\end{eqnarray}
where $p^+$ is arbitrary, and $q^\prime$ is the momentum of
the final (on-shell) photon.
For simplicity we choose $p^+=1$,
and we have $q^2=-q_\bot^2=-Q^2$. 
Then the $F_{\eta_c \gamma}$ is given by 
\begin{eqnarray}
F_{\eta_c \gamma}(Q^2)=\frac{\Gamma^+}{-i e (\epsilon_\bot \times q_\bot)},
\end{eqnarray}
where $\epsilon=(0,0,\epsilon_\bot)$ and $\epsilon_\bot \cdot q_\bot=0$
is chosen.

The contribution coming from the ordinary helicity 
components ($\lambda_1+\lambda_2=0$)
turns out to be 
\begin{eqnarray}
F^{(\lambda_1+\lambda_2=0)}_{\eta_c \gamma}(Q^2)
&=&\frac{\sqrt{n_c}e_c^2}{i(\epsilon_\bot \times
q_\bot)}\int_0^1[{\rm d}x] \int_0^\infty \frac{{\rm d}^2 k_\bot}{16 \pi^3}
\frac{m_c}{\sqrt{m_c^2+k_\bot^2}}\psi(x_i,k_\bot) \nonumber \\
&~~~~\times& \left[\frac{\bar v_\downarrow (x_2,-k_\bot)} 
{\sqrt{x_2}}\rlap /\epsilon
\frac{u_\uparrow (x_1,k_\bot+q_\bot)}{\sqrt{x_1}} 
\frac{\bar u_\uparrow (x_1,k_\bot+q_\bot)}{\sqrt{x_1}} 
\gamma^+
\frac{u_\uparrow (x_1,k_\bot)}{\sqrt{x_1}} 
\frac{1}{D} + (1 \leftrightarrow 2) \right],
\label{spinor}
\end{eqnarray}
where $[{\rm d} x]= {\rm d}x_1 {\rm d}x_2 \delta(1-x_1-x_2),$
$e_c$ is the $c$-quark charge in unit of $e$,
and  $D$ is the ``energy-denominator",
\begin{eqnarray}
D&=&q_\bot^2-\frac{(q_\bot+k_\bot)^2+m_c^2}{x_1}
-\frac{k_\bot^2+m_c^2}{x_2} \nonumber\\
&=&-\frac{(x_2 q_\bot+k_\bot)^2+m_c^2}{x_1 x_2}
\end{eqnarray}
Being different from
the case of the light meson such as $\pi, \eta$ and $\eta^\prime$,
the present of the large quark mass ($m_c\simeq 1.5$ GeV)
always prevent $1/D$ from the
singular point $D\ra 0$, {\it i.e.} the partons in the intermediate state
are always far off energy-shell. This means that even at the low $Q^2$
region, the LCPQCD calculation may be still available.
Employing the LCPQCD calculation rules
Eq.~(\ref{spinor}) becomes \cite{Cao,LCPQCD},
\begin{eqnarray}
F^{(\lambda_1+\lambda_2=0)}_{\eta_c \gamma}(Q^2)
&=&2\sqrt{2} \sqrt{n_c} e_c^2 \int_0^1 [{\rm d}x]
\int \frac{{\rm d}^2 k_\bot}{16 \pi^3}
\frac{m_c}{\sqrt{m_c^2+k_\bot^2}}\psi(x_i,k_\bot) \nonumber \\
&~~~~\times&\left[
\frac{q_\bot \cdot (x_2 q_\bot + k_\bot)}
{q_\bot^2 [(x_2 q_\bot + k_\bot)^2+m_c^2]} +(1 \leftrightarrow 2)
\right].
\label{fh0}
\end{eqnarray}

Similarly, one can obtain
the contribution coming from the higher helicity components,
\begin{eqnarray}
F^{(\lambda_1+\lambda_2=\pm 1)}_{\eta_c \gamma}(Q^2)&=&
\frac{\sqrt{n_c}e_c^2}{i(\epsilon_\bot \times
q_\bot)}\int_0^1[{\rm d}x] \int_0^\infty \frac{{\rm d}^2 k_\bot}{16 \pi^3}
\frac{m_c}{\sqrt{m_c^2+k_\bot^2}}\psi(x_i,k_\bot) \nonumber \\
&~~~~\times& \left[\frac{\bar v_\uparrow (x_2,-k_\bot)} 
{\sqrt{x_2}}\rlap /\epsilon
\frac{u_\uparrow (x_1,k_\bot+q_\bot)}{\sqrt{x_1}} 
\frac{\bar u_\uparrow (x_1,k_\bot+q_\bot)}{\sqrt{x_1}} 
\gamma^+
\frac{u_\uparrow (x_1,k_\bot)}{\sqrt{x_1}} 
\frac{1}{D} + (1 \leftrightarrow 2) \right] \nonumber \\
&=& 2\sqrt{2} \sqrt{n_c} e_c^2 \int_0^1 [{\rm d}x]
\int \frac{{\rm d}^2 k_\bot}{16 \pi^3}
\frac{m_c}{\sqrt{m_c^2+k_\bot^2}}\psi(x_i,k_\bot) \nonumber \\
&~~~~\times&\left[\frac{q_\bot \cdot k_\bot}
{q_\bot^2 [(x_2 q_\bot + k_\bot)^2+m_c^2]} +(1 \leftrightarrow 2) \right].
\label{fh1}
\end{eqnarray}
Once again,
a non-zero quark mass, $m_c$ plays an important role in the calculation
of $F^{(\pm 1)}_{\eta_c \gamma}$, since in the $m_c \ra 0$ limit the matrix
$\bar v_{\uparrow (\downarrow)} (x_2, -k_\bot)\rlap /\epsilon
u_{\uparrow (\downarrow)}(x_1, q_\bot +k_\bot)$
will goes to zero.
Therefore, for the light meson such as $\pi$, $\eta$ and $\eta^\prime$
neglecting the contributions coming from higher helicity components
should be a good approximation.
Combining this matrix with the coefficients $C_0(x, k_\bot,\uparrow,\uparrow)$
and $C_0(x, k_\bot,\downarrow,\downarrow)$, one arrives the second
expression in Eq. (\ref{fh1}).
The full result is obtained by summing up
the contributions from the ordinary helicity components (Eq. (\ref{fh0}))
and that from the higher helicity components (Eq. (\ref{fh1})),
\bea
F_{\eta_c \gamma}(Q^2)=F^{(\lambda_1+\lambda_2=0)}_{\eta_c \gamma}(Q^2)
+F^{(\lambda_1+\lambda_2=\pm 1)}_{\eta_c \gamma}(Q^2).
\label{fha}
\eea

Neglecting $k_\bot$ and $m_c$ relative to $x_2 q_\bot$
in Eqs. (\ref{fh0}) and (\ref{fh1}), and employing the
asymptotic form distribution
amplitude\footnote{Any meson distribution amplitude should evolve into 
the asymptotic form in the $Q^2 \rightarrow \infty$ limit.}
\bea
\phi(x)=\sqrt{3/2} \ f_{\eta_c} x_1 x_2,
\eea
where $f_{\eta_c}$ is the decay constant,
one can obtain the asymptotic prediction for the $\eta_c$-$\gamma$ transition
form factor,
\bea
F_{\eta_c\gamma}(Q^2\ra \infty)=\frac{8 f_{\eta_c}}{3 Q^2}.
\label{fas}
\eea
Corrections to the asymptotic prediction (Eq. (\ref{fas}))
come from $c$-quark mass, the $k_\bot$-dependence and the
higher helicity components (see Eqs. (\ref{fh0}),
(\ref{fh1}) and (\ref{fha})).
All of these corrections are suppressed by the factor $1/Q^2$
at the large-$Q^2$ region.
But in the present experimentally available energy region,
these contributions may be important and should be taken into account.

In order to study the $c$-quark mass effect,
one may first neglect the $k_\bot$-dependence in the hard-scattering
amplitude of Eq. (\ref{fh0}),
then one can obtain,
\bea
F_{\eta_c\gamma}(Q^2)=
2\sqrt{2} \sqrt{n_c} e_c^2 \int_0^1 [{\rm d}x]\phi(x)
\left[\frac{x_2 }{(x_2 q_\bot)^2 +m_c^2}+(1 \leftrightarrow 2)\right],
\label{fmc}
\eea
where $\phi(x)$ is the distribution amplitude of the $\eta_c$-meson,
\bea
\phi(x)=\int_0^\infty \frac{{\rm d}^2 k_\bot}{16 \pi^3}
\frac{m_c}{\sqrt{m_c^2+k_\bot^2}}\psi(x, k_\bot).
\eea
Because of the $c$-quark mass being large, Eq.~(\ref{fmc})
will approach to the asymptotic prediction (Eq.~(\ref{fas}))
in a rather slow way,
that is, the corrections coming from $c$-quark mass effect are large
in a rather broad energy region.
The effects of the $k_\bot$-dependence and higher helicity components
can be studied by comparing the results obtained from
Eqs. (\ref{fh0}), (\ref{fh1}), (\ref{fha}) and (\ref{fmc}).
Also, it is interesting to notice that the correction
coming from the higher helicity components
is the same as that from the $k_\bot$-dependence
in the ordinary helicity components (The right hand side in
Eq. (\ref{fh1}) is the same as the second term in the
right hand side of Eq. (\ref{fh0})).
In the low and medium $Q^2$ region,
these corrections may provide sizable contributions which
should be taken into account.

We point out that the above analysis for
$F_{\eta_c\gamma}$ is applicable to the form factor
$F_{\eta_b\gamma}$ with the physics quantities corresponding
to the $c$ quark ($e_c$ and $m_c$) 
and decay constant $f_{\eta_c}$ being replaced by
the ones corresponding to the $b$ quark ($e_b$ and $m_b$) 
and $f_{\eta_b}$, respectively.
The differences resulting from the $b$-quark being much heavier
than the $c$-quark are as follows:
First, the modification coming from $b$-quark mass effect become
much severer,
{\it i.e.} the perturbative calculation with $m_b$ effect
being respected approaches to the asymptotic prediction
more slowly.
Second, the corrections coming from the transverse momentum
dependence and the higher helicity components of the light-cone
wave function may become rather mild because the
$b$-quark mass is much larger than the parton's average
transverse momentum.

\section{Numerical calculations}

We employ the Brodsky-Huang-Lepage (BHL) model \cite{BHL}
for the $\eta_c$ ($\eta_b$) meson light-cone wave functions,
\begin{eqnarray}
\psi^{BHL}(x,k_\bot)= A \, \rm {exp}\left[ -\frac{k_\bot^2+m^2}
{8 \beta^2 x (1-x)} \right ].
\label{model1}
\end{eqnarray}
In this model, the light-cone wave function is obtained from
the instant-form wave function by demanding the off-shell energies
being equal in the two reference frames.
The parameters $A$ and $\beta$ are determined by the following two
constraints:
\bea
\int_0^1[dx]\int\frac{d^2{k_\bot}}{16\pi^3}
\frac{m_q}{\sqrt{m_q^2+k^2_\bot}}\psi(x,k_\bot)
=\frac{f_{\eta_q}}{\sqrt{6}},
\eea
\bea
\int_0^1[dx]\int\frac{d^2{k_\bot}}{16\pi^3}
\left | \psi(x,k_\bot) \right |^2
=P_{q{\bar q}/\eta_q}
\eea
where $f_{\eta_q}$ ($q=c,b$) is the decay constant of the
$\eta_c$ ($\eta_b$ ) meson corresponding to
$f_\pi=131$ MeV, and $P_{q{\bar q}/\eta_q}$ is the probability of
finding $\left | c {\bar c} \right. \rangle$ 
$\left( \left |b {\bar b} \right. \rangle \right)$ Fock state
in the $\eta_c$ ($\eta_b$) meson.
Because of the lack of experimental information, one often
evaluates $f_{\eta_q}$ through various theoretical approaches.
Employing the Van Royen-Weisskopf formula \cite{Royen}
for the decay constant\footnote{The decay constants of the
pseudoscalar and vector mesons are defined by
$\langle 0 | {\bar Q}\gamma^\mu\gamma_5 Q^\prime | M_P({\bf K}) \rangle
=f_P K^\mu$ and
$\langle 0 | {\bar Q}\gamma^\mu Q^\prime | M_V({\bf K},\varepsilon) \rangle
=f_V m_V \varepsilon^\mu$, respectively, where $\varepsilon$ is the
polarization vector of the vector meson, and $K$ is the meson momentum.}
\bea
f_M=\sqrt{\frac{12}{m_M}} \left | \psi_M(0) \right |
\eea
where $m_M$ and $\psi_M(0)$ are the mass and wave function at the
origin of the meson respectively,
one can obtain that the decay constant of the pesudoscalar meson
is almost the same as that of the vector meson, {\it i.e.}
$f_P=f_V$. Although the hyperfine splitting Hamiltonian
may destroy this relation \cite{Ahmady},
the consideration of the difference
coming from the mock meson spin structure may rescue it \cite{Hwang}.
Hence, we adopt \cite{Hwang,Chao}
\bea
f_{\eta_c}\simeq f_{J/\psi}\simeq 420 \ {\rm MeV}, \ \
f_{\eta_b}\simeq f_{\Upsilon}\simeq 705 \ {\rm MeV}.
\eea
As well known, with
the increasing of the constitute quark mass the valence Fock state
occupies a bigger fraction in the hadron, and in the nonrelativistic
limit the probability of finding the valence Fock state is going to
approach unity. So one can expect
$P_{q{\bar q}/\eta_q}=0.8 \sim 1.0$. Our calculation shows that
the prediction for
the $\eta_c$ ($\eta_b$) transition form factor,
$F_{\eta_c\gamma}$ ($F_{\eta_b\gamma}$)
are not sensitive to the value of $P_{c{\bar c}/\eta_c}$
($P_{b{\bar b}/\eta_b}$) \cite{Feldmann}.
So we may take
\bea
P_{c{\bar c}/\eta_c}=0.8, \ \
P_{b{\bar b}/\eta_b}=1.0.
\eea
From the above constrains, one can obtain the parameters in the
wave functions,
\bea
  A=54.44 \ {\rm MeV}^{-1},  & \ \ \ \beta=0.994 \ {\rm MeV}
& \ \ \ {\rm for} \ \  \eta_c, \\
  A=4146 \ {\rm MeV}^{-1},   & \ \ \ \beta=1.507 \ {\rm MeV}
& \ \ \ {\rm for} \ \ \eta_b.
\eea
The average transverse momenta of the quark in the mesons defined by
$\langle k_\bot \rangle=\sqrt{\langle {\bf k}^2_\bot \rangle}$ with
\bea
\langle {\bf k}^2_\bot \rangle
=\frac{1}{P_{q{\bar q}/\eta_q}} \int_0^1[dx]\int\frac{d^2{k_\bot}}{16\pi^3}
|{\bf k}_\bot|^2 \psi(x,k_\bot)
\eea
turn out to be $950 \ {\rm MeV}$ and $1.48 \ {\rm GeV}$ for the
$\eta_c$ and $\eta_b$, respectively.

We present our numerical results for $F_{\eta_c\gamma}$ in figure 2.
The dash-dotted line is the asymptotic prediction (Eq. (\ref{fas})).
The solid curve is obtained by respecting the $m_c$ effect
but neglecting the corrections from the $k_\bot$-dependence
and the higher helicity components (Eq. (\ref{fmc})).
The dashed curve is obtain by taking into account $m_c$ effect
and the $k_\bot$-dependence
in the ordinary light cone wave function
but neglecting the contributions form
the higher helicity components (Eq. (\ref{fh0})).
Considering all of these corrections gives
the dotted curve (Eq. (\ref{fha})).
In the $Q^2\rightarrow \infty$ limit, all of these calculations
approach the asymptotic prediction.
But, because of the $c$-quark being heavy,
taking into account the quark mass effect
significantly modifies the perturbative prediction
in a rather broad energy region.
At $Q^2 \simeq 10 \ {\rm GeV}^2$, the result obtained by including
the $c$-quark mass effect is only about $1/3$
of the asymptotic prediction for the $F_{\eta_c\gamma}$.
At $Q^2 \simeq 100 \ {\rm GeV}^2$ the ratio is about $70\%$.
Also it can be found that in the energy region of
$Q^2 \leq 10 \, {\rm GeV}^2$ where the 
the present experiments are able to approach,
the parton's transverse momentum and
higher helicity components bring sizable corrections
to the prediction of $F_{\eta_c\gamma}$.

The numerical results for $F_{\eta_b\gamma}$ are given in figure 3.
The curve explanations are similar as that in figure 2.
It can be found that the modification resulting from the $b$-quark
mass effect is much severer than that in the case of $F_{\eta_c\gamma}$,
because $b$-quark is heavier than the $c$-quark.
At $Q^2 \simeq 10 \ {\rm GeV}^2$, the result obtained by including
the $b$-quark mass effect is only about $1/15$
of the asymptotic prediction for the $F_{\eta_b\gamma}$.
At $Q^2 \simeq 100 \ {\rm GeV}^2$ the ratio is about $30\%$.
On the other hand, the corrections coming form
the parton's transverse momentum and higher helicity components
are negligible in the calculation of $F_{\eta_b\gamma}$
since the $b$-quark mass,
$m_b$ is much heavy than the parton's average transverse momentum
in the $\eta_b$ meson.
One can expect that LEP2 may examine all of these theoretical
predictions in the near future.

\section{Summary}
In summary, the meson photon transition form factors $F_{P\gamma}(Q^2)$
($P$ being $\pi^0,\eta,\eta^\prime,\eta_c,\eta_b$ \ ...)
extracted from the two photon collision are the simple exclusive processes
which can provides a rather simple and rigorous way to the
test of QCD and the determination of the meson wave function
(non-perturbative physics). Many experimental collaboration such as
TPC/Two-Gamma, CELLO, CLEO and L3 {\it etc.} have studied these processes.
A measurement for the $F_{\eta_c\gamma}(Q^2)$ is very likely to be
feasible in LEP2.
In this note, we analyze the $\eta_c$- and $\eta_b$-photon
transition form factors
in the light-cone perturbative theory with the quark mass effect,
the parton's transverse momentum dependence and the higher helicity
components of the light cone wave function are respected.
It is pointed out that due to $c$- ($b$-) quark being heavy,
considering the quark mass effect
brings significant modifications to
the perturbative predictions in a rather broad energy region.
This effect is much severer for the $F_{\eta_b\gamma}$ than
that for the $F_{\eta_c\gamma}$ because of the $b$-quark being
heavier than $c$-quark.
Also it is found that, for the $F_{\eta_c\gamma}$,
the parton's transverse momentum and
higher helicity components bring sizable corrections
in the present experimentally accessible
energy region ($Q^2 \leq 10\sim 20 \ {\rm GeV}^2$),
while these corrections are negligible in the perturbative calculation
of $F_{\eta_b\gamma}$.
We conclude that the coming $e^+ e^-$ collider LEP2
will provide the opportunity to examine all of these
theoretical predictions.

\vskip 0.5cm
\centerline{{\bf Acknowledgments}}
This work partially supported by the
Postdoc Science Foundation of China and the National Natural
Science Foundation of China.


\vskip 1cm
\section*{Figure Captions}
\begin{description}
\item
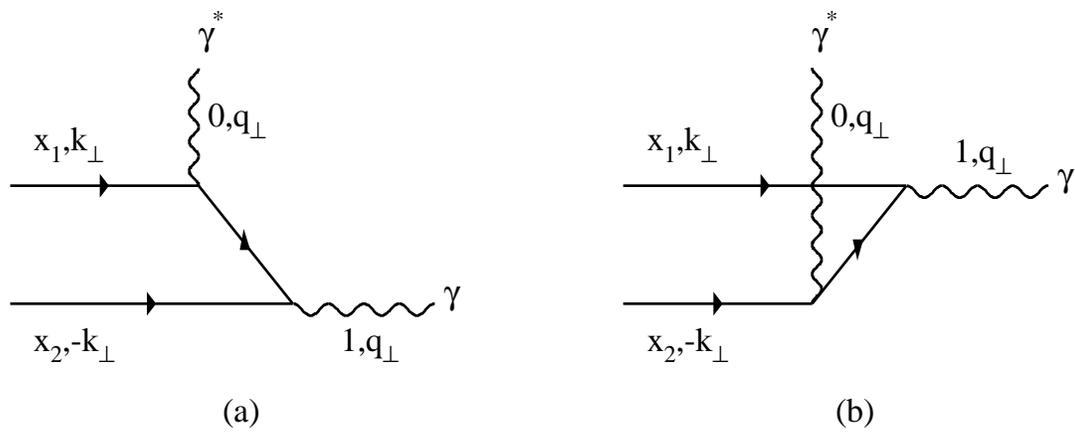
{Fig.~1} The lowest order diagrams contributing to $F_{\eta_c \gamma}$ 
and $F_{\eta_b \gamma}$ 
in the light-cone perturbative QCD. The momenta are expressed in the
light-cone variables $(+,\bot)$.
\item
{Fig.~2(a)} The $\eta_c$-$\gamma $ transition form factor given in
$Q^2F_{\eta_c\gamma}(Q^2)$.
\item
{Fig.~2(b)} The $\eta_c$-$\gamma $ transition form factor given in
$F_{\eta_c\gamma}(Q^2)$.
\item
{Fig.~3(a)} The $\eta_b$-$\gamma $ transition form factor given in
$Q^2F_{\eta_b\gamma}(Q^2)$.
\item
{Fig.~3(b)} The $\eta_b$-$\gamma $ transition form factor given in
$F_{\eta_b\gamma}(Q^2)$.

\end{description}

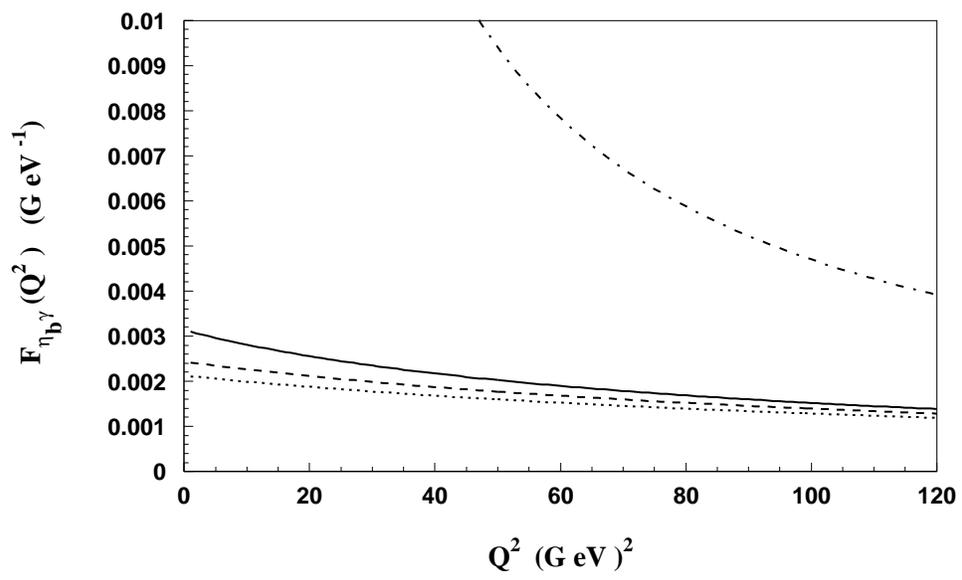
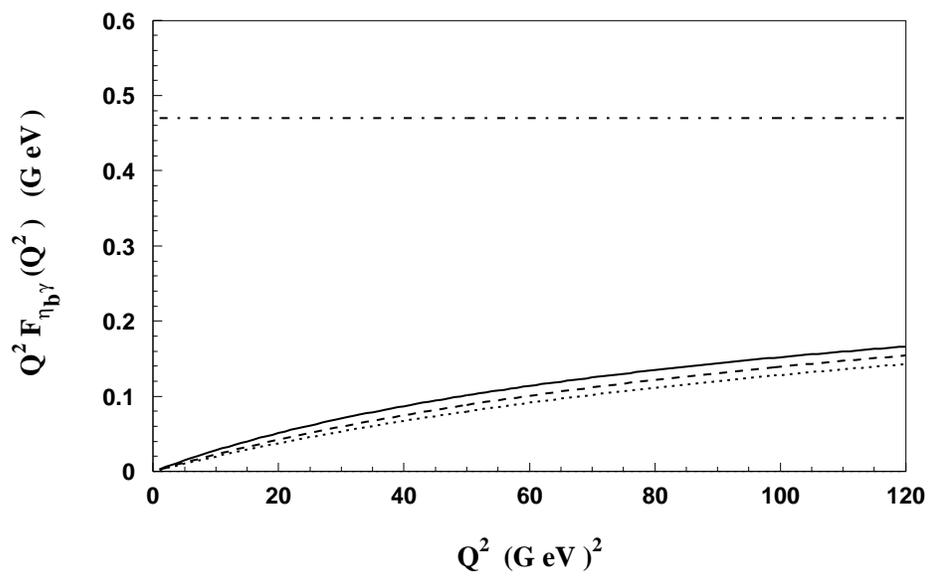
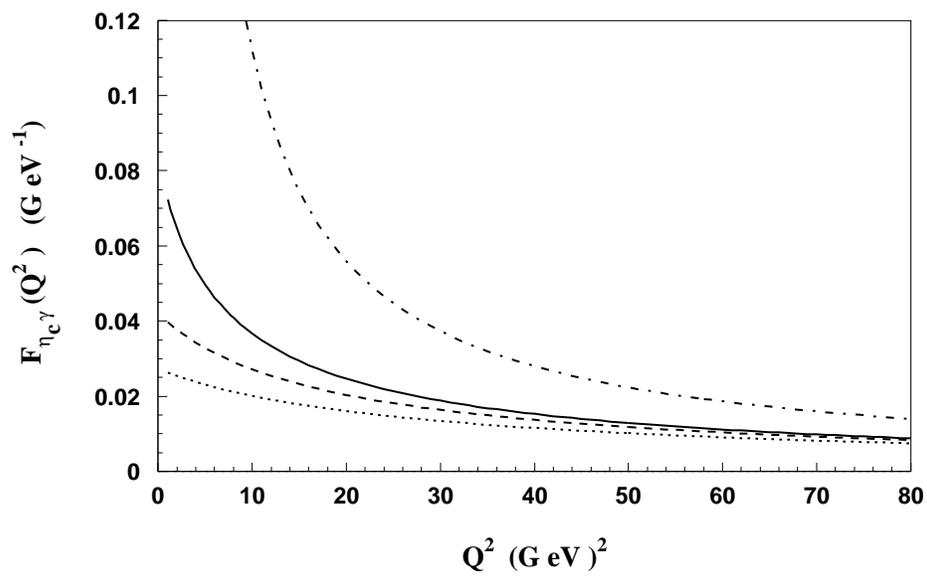
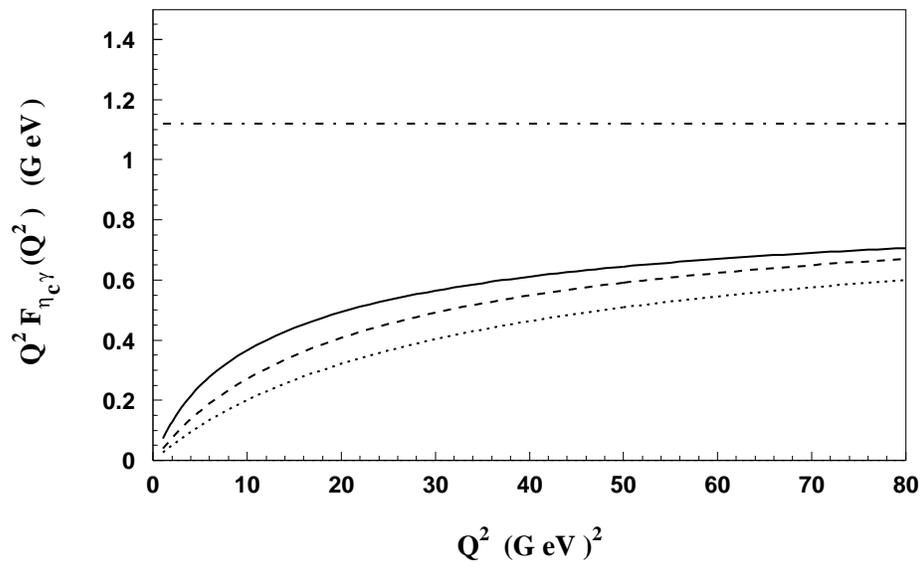
\end{document}